\documentstyle[12pt,worldsci,epsf,rotate]{article}
\begin{document}
\title{NEWS FROM THE LATTICE}
\author{C.~Morningstar\\
 {\em University of California at San Diego,
  La Jolla, California 92093-0319}}
\maketitle
\setlength{\baselineskip}{2.6ex}

\vspace{0.7cm}
\begin{abstract}

A summary of some recent results from lattice simulations
is presented.  These include first calculations of the
strangeness magnetic moment of the nucleon, three new studies of
the gluon propagator, flux tube attraction in $U(1)$ gauge
theory, and the static-quark potential and its gluonic
excitations.

\end{abstract}
\vspace{0.7cm}

\section{Introduction}

Lattice simulations afford a means of studying confinement from
first principles.  Faster computers and more efficient simulation
techniques (duality transformations, improved actions, anisotropic
lattices) are allowing
access to physics in previously unexplored territories: the
strangeness content of the nucleon, the infrared region of the
Landau-gauge gluon propagator, the long flux in $U(1)$, the rich
glueball spectrum, and the static-quark potential and its gluon
excitations for quark-antiquark separations as large as 4 fm.  
In this talk, the new lattice results presented at this conference
are summarized.

\section{Strangeness magnetic moment of the nucleon}

The strangeness content of the nucleon is currently of much interest.
A significant $\bar{s}s$ content in the nucleon can resolve the
long-standing discrepancy between the pion-nucleon sigma term
extracted from low-energy pion-nucleon scattering and that from
the octet baryon masses.  A large and negative strange-quark 
polarization has emerged from spin structure function
studies at CERN and SLAC, combined with neutron and hyperon $\beta$
decays.  Substantial contributions from strange quarks in the
axial-vector and scalar current matrix elements would seem to imply
similar such contributions to the vector current, but this would
jeopardize the $SU(6)$ prediction of the neutron to proton magnetic
moment ratio, which lends credence to the valence quark
picture.  The SAMPLE collaboration has recently measured the
neutral weak magnetic form factor in elastic parity-violating electron
scattering at backward angles, and obtained the strangeness magnetic
form factor by subtracting out the nucleon magnetic form
factor.  Elastic $ep$ and $e\,{}^{4}\!He$ parity-violation experiments
are currently planned at TJNAF to measure the asymmetry at forward
angles to extract the strangeness electric mean-square radius.

Williams presented results\cite{smagmom} from the first lattice calculation
of the strangeness magnetic $G^s_M(q^2)$ and electric $G^s_E(q^2)$ form
factors, where $q$ is the four-momentum transfer.  These form factors
were obtained from ratios of two- and three-point correlation functions
involving the vector current.  The simulations were carried out on a
$16^3\times 24$ lattice with an inverse spacing $a^{-1}=1.72(4)$ GeV
and were possible only because of recent developments in $Z_2$ noise
techniques for quark propagators.  The calculations were done in the
quenched approximation; contributions due to sea quarks were explicitly
included at one fermion-loop order (the disconnected insertion
diagram).  The connected insertion diagram yields the contributions
from the valence quarks and their $Z$ graphs.  Three values for the
Wilson hopping parameter were used, corresponding to quark masses of
120, 200, and 360 MeV.  

\begin{figure}[t]
\begin{center}
\leavevmode
\epsfxsize=5.8in\epsfbox[86 366 466 718]{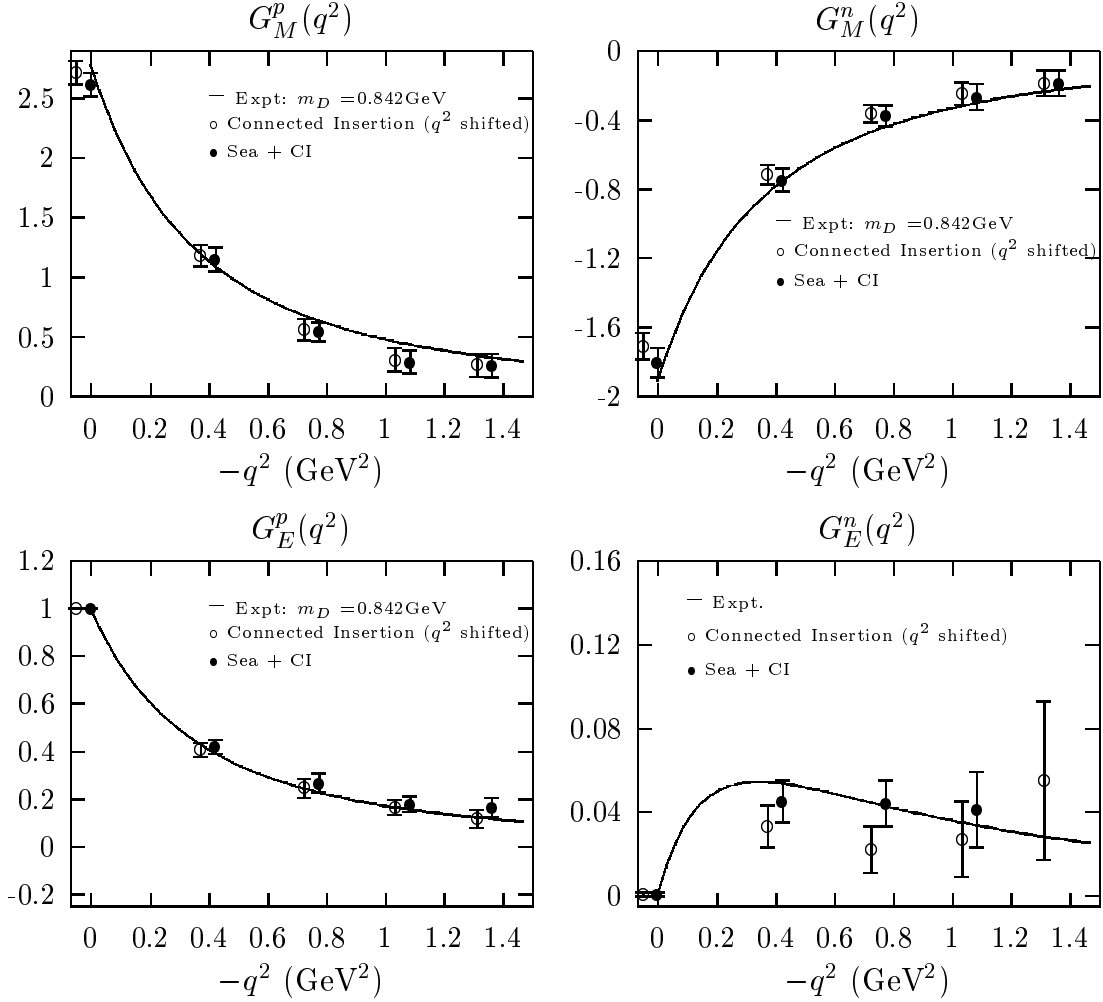}
\end{center}
\vspace{-8mm}
\caption[figsmom]{
  Magnetic $G_M^N(q^2)$ and electric $G_E^N(q^2)$ form factors for the
  proton ($N=p$) and the neutron ($N=n$), where $q$ is the four-momentum
  transfer.  The open circles indicate results from the connected
  insertion only, while the full results are shown as solid circles.
  The open circles are shifted slighted to the left to avoid overlap.
  The solid line is the fit to experiment.
  }
\label{fig:smagmom}
\end{figure}

Results for $G^s_M(q^2)$ and $G^s_E(q^2)$ were
obtained at four nonzero values of $q^2$.  Extrapolating to the
limit $q^2\!\rightarrow\!0$ using a single pole form yielded
$G_M^s(0) = - 0.36 \pm 0.20 $ and a value for $G_E^s(0)$ consistent
with zero.  The electric mean-square radius was 
$\langle r_s^2 \rangle_E = - 0.061 \pm 0.003\,{\rm fm}^2$.
For the $u$ and $d$ quarks in the
disconnected insertion diagram, $G_{M,{\rm dis}}^{u/d}(0) 
= -0.65 \pm 0.30$ was found.  Adding all of the sea quark contributions
from the disconnected insertion diagram gives $\mu_{dis} = 
(2/3 G_{M,{\rm dis}}^u(0) -1/3 G_{M,{\rm dis}}^d(0) -1/3 G_M^s(0))
\mu_N = -0.097 \pm 0.037 \mu_N$, which, when combined with the connected
insertion diagram, brings the ratio of neutron to proton magnetic
moments $\mu_n/\mu_p = -0.68 \pm 0.04$ into agreement
with the experimental value of 0.685.  The accidental cancellation
between the disconnected insertion diagram and the rest-frame $Z$-graph
component of the connected insertion diagram leaves the nonrelativistic
valence contribution dominant, which explains why the
$SU(6)$ valence quark picture works well for the $\mu_n/\mu_p$ ratio;
$SU(6)$ fails in the axial-vector and scalar cases because this
cancellation does not occur.  The results for the form
factors are shown in Fig.~\ref{fig:smagmom}.  Contributions from
the strange quarks are not particularly small, but they are largely
cancelled by the $u$ and $d$ quark contributions, resulting in a
small overall effect to the form factors (the effect is greatest
in the nucleon electric form factor).  Issues related to the
quenched approximation, chiral extrapolations, finite lattice-spacing
artifacts, and finite volume errors must still be explored.

\section{Gluon propagator}

The connection between confinement and the gauge-dependent gluon
propagator is unclear.  However,
knowledge of the infrared behaviour of the gluon propagator 
is important for various approaches to modelling
confinement.  Some studies based on Dyson--Schwinger equations claim that an
infrared enhanced (such as $1/q^4$) propagator is required for confinement;
others assume that a dynamically generated gluon mass leads to an
infrared finite behaviour.  Lattice simulations can tell us the
true infrared behaviour of the gluon propagator from first principles.
However, large finite-volume effects have prevented previous lattice
studies from accessing the momentum region where the relevant
nonperturbative behaviour is expected.  At this conference, three
groups reported on new lattice studies of the gluon propagator;
two worked in the Landau gauge and one used the maximally-abelian gauge
in $SU(2)$ with residual $U(1)$ Landau gauge fixing. 

\begin{figure}[t]
\begin{center}
\leavevmode
\setlength{\unitlength}{0.1in}
\begin{picture}(58,20)
\put(24,17){{\Large (a)}}
\put(36,17){{\Large (b)}}
\epsfysize=2.8in
\rotate{\vbox{\epsfbox[55 63 530 734]{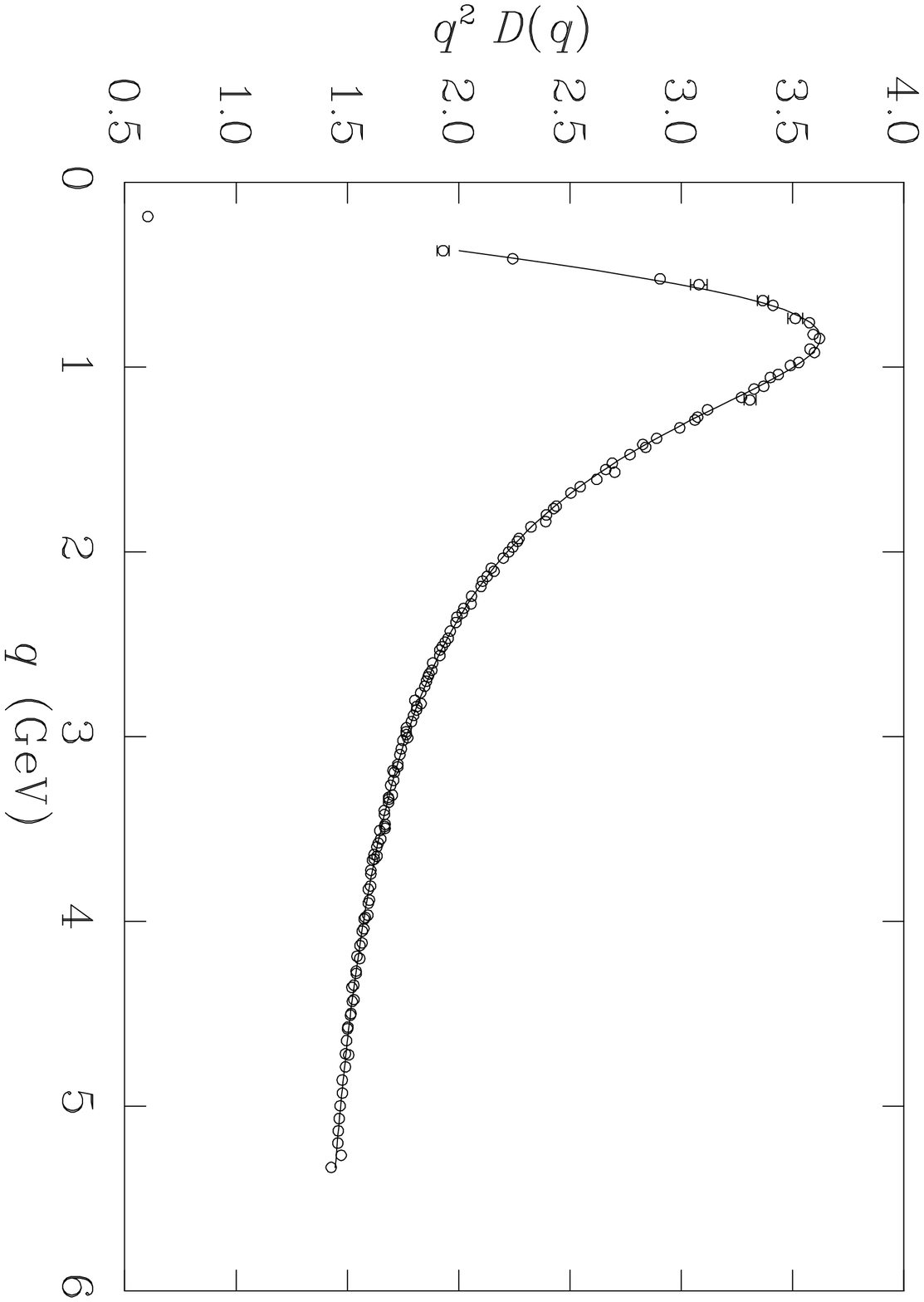}}}
\epsfxsize=2.9in\epsfbox[50 0 288 165]{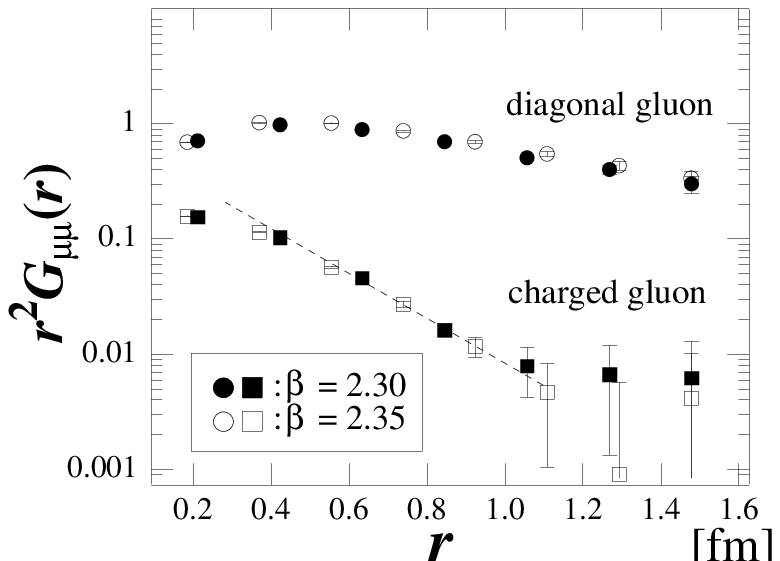}
\end{picture}
\end{center}
\vspace{-5mm}
\caption[figglue]{(a) The Landau-gauge gluon propagator in momentum
  space multiplied by $q^2$, with nearby points 
  averaged\protect\cite{gluonpropA}. The line is a best fit using the form
  given in  Eq.~(\protect{\ref{eq:gluemodel}}). The scale is set using
  the string tension. (b)  The $SU(2)$ gluon propagator\cite{gluonpropB}
  in the maximally-abelian gauge with residual $U(1)$ Landau gauge fixing.
  The colour diagonal and off-diagonal components are shown.
  }
\label{fig:glueres}
\end{figure}

Williams presented results,\cite{gluonpropA}
shown in Fig.~\ref{fig:glueres}(a), revealing new structure
in the gluon propagator for momenta as small as 0.4 GeV.
Three separate simulations were done in order to control systematic errors
from finite lattice spacing and finite volume.  They obtained results
in the quenched approximation on a large ($32^3\times 64$) lattice for
a lattice spacing near 0.1 fm, which enabled them to probe momenta
as small as 400 MeV.  Gribov copies were ignored.  They also verified
the tensor structure of the propagator, given in the continuum by
\begin{equation}
D_{\mu\nu}^{ab}(q) =
\delta^{ab}\left(\delta_{\mu\nu}-\frac{q_{\mu}q_{\nu}}{q^2}\right)D(q^2),
\end{equation}
where $q$ is the four-momentum of the gluon, $a$ and $b$
are colour indices, and $\mu$ and $\nu$ are the space-time indices.
One of their main findings was to rule out a $1/q^4$ infrared enhancement
(for the case when quark-antiquark pair creation is neglected).
They fit several phenomenological functions to their data, but most 
resulted in very large $\chi^2$ values.  In particular, the Gribov,
Stingl, Marenzoni, and Cornwall forms were all shown to fail.
One of the few functions
to provide a satisfactory description of the data over a wide
range of momenta was found empirically to be
\begin{equation}
D(q^2) =  Z\left(\frac{A}{(q^2)^{1+\alpha}+(M_{\rm IR}^2)^{1+\alpha}} +
\frac{1}{q^2+M_{\rm UV}^2}\left(\frac{1}{2}
\ln\frac{q^2+M_{\rm UV}^2}{M_{\rm UV}^2}\right)^{-d_D}\right), 
\label{eq:gluemodel}
\end{equation}
where $Z=1.78^{+45}_{-20}$, $A=0.49^{+17}_{-6}$, $aM_{\rm UV}=0.20^{+37}_{-19}$,
$aM_{\rm IR}=0.43^{+5}_{-1}$, and $\alpha=0.95^{+7}_{-5}$ were the fit
parameters, and $a^{-1}=1.885$ GeV and $d_D=13/44$.  This function does
incorporate some ultraviolet information from perturbation theory.
In the future, these
authors intend to use an improved-gauge field action and increase
the lattice volume in order to probe deeper into the infrared.

Results for the $SU(2)$ gluon propagator in the maximally-abelian
gauge (with residual $U(1)$ Landau gauge fixing) were presented
by Suganuma\cite{gluonpropB} and are shown in Fig.~\ref{fig:glueres}(b).
Two simulations on a $12^3 \times 24$ lattice were done, but the
lattice spacings were too similar to allow any serious study of
systematic errors from finite spacing and finite volume.  Their focus
was to demonstrate that the colour off-diagonal component has a Yukawa
fall-off corresponding to a large mass so that long-range physics
($r>0.35$ fm) must be dominated by the colour diagonal component.
This, they claimed, is the origin of abelian dominance in
nonperturbative QCD.  The dependence of the off-diagonal gluon's
effective mass on the residual $U(1)$ gauge fixing was not studied.
It should be noted here that abelian dominance of
nonperturbative QCD has not yet been convincingly established, and
that confinement is much more than the area law of the Wilson loop.

Nakajima and Furui\cite{naka} presented preliminary results
from not only quenched
simulations, but also from Langevin simulations including dynamical fermions
(for three values of the Wilson hopping parameter $\kappa=0.1$, 0.15, and 0.2).
Smeared gauge fixing was used to resolve Gribov ambiguities.  Their
lattices were very small ($4^3\times 8)$ and much more work is needed
to control systematic errors before any conclusions can be drawn on the
role of quark-antiquark pair creation in the gluon propagator.

\section{Dual lattice simulations of $U(1)$ flux tubes}

The study of confinement in the strong-coupling phase of compact
$U(1)$ in four dimensions can be viewed as a stepping stone on the
way to understanding confinement in QCD.  Confinement in the $U(1)$
theory is driven by magnetic monopoles whose currents squeeze
the electric field into flux tubes.  A dual transformation of the
$U(1)$ path integral exists so that simulations can be carried out
much more efficiently in the dual theory, which corresponds to a
certain limit of a dual Higgs model.
Results obtained in the dually-transformed theory were
presented\cite{dualsim} at this conference, along with some first
computations in the dual Abelian Higgs model.  This study focussed
on the question of whether the vacuum corresponds to a type-I or
a type-II superconductor, and the role of quantum fluctuations of
the dual degrees of freedom were investigated.

\begin{figure}[t]
\begin{center}
\leavevmode
\setlength{\unitlength}{0.1in}
\begin{picture}(58,16)
\put(4,13){{\Large (a)}}
\put(33,13){{\Large (b)}}
\epsfxsize=2.9in\epsfbox[80 432 542 692]{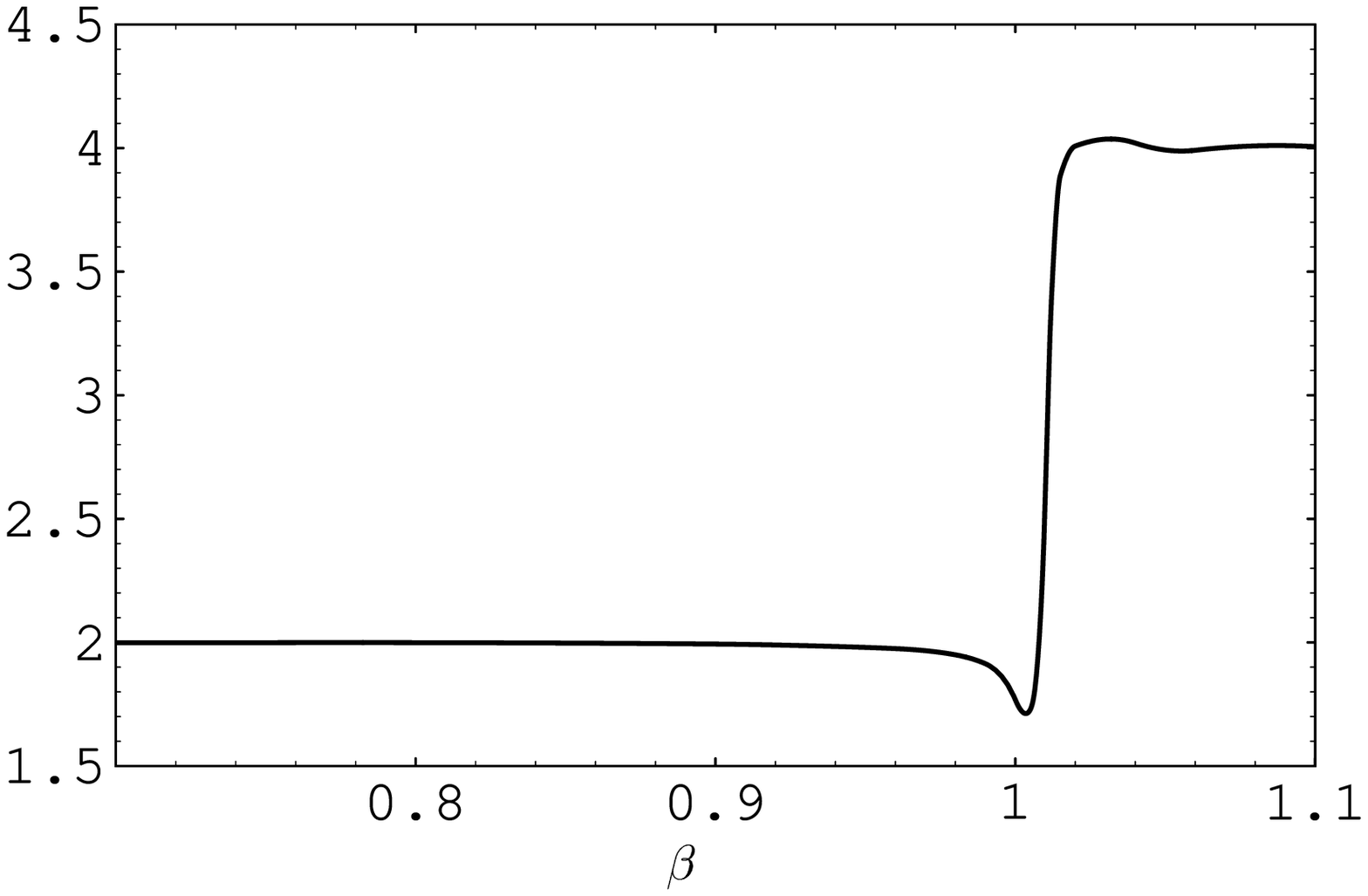}
\epsfxsize=2.9in\epsfbox[120 530 452 722]{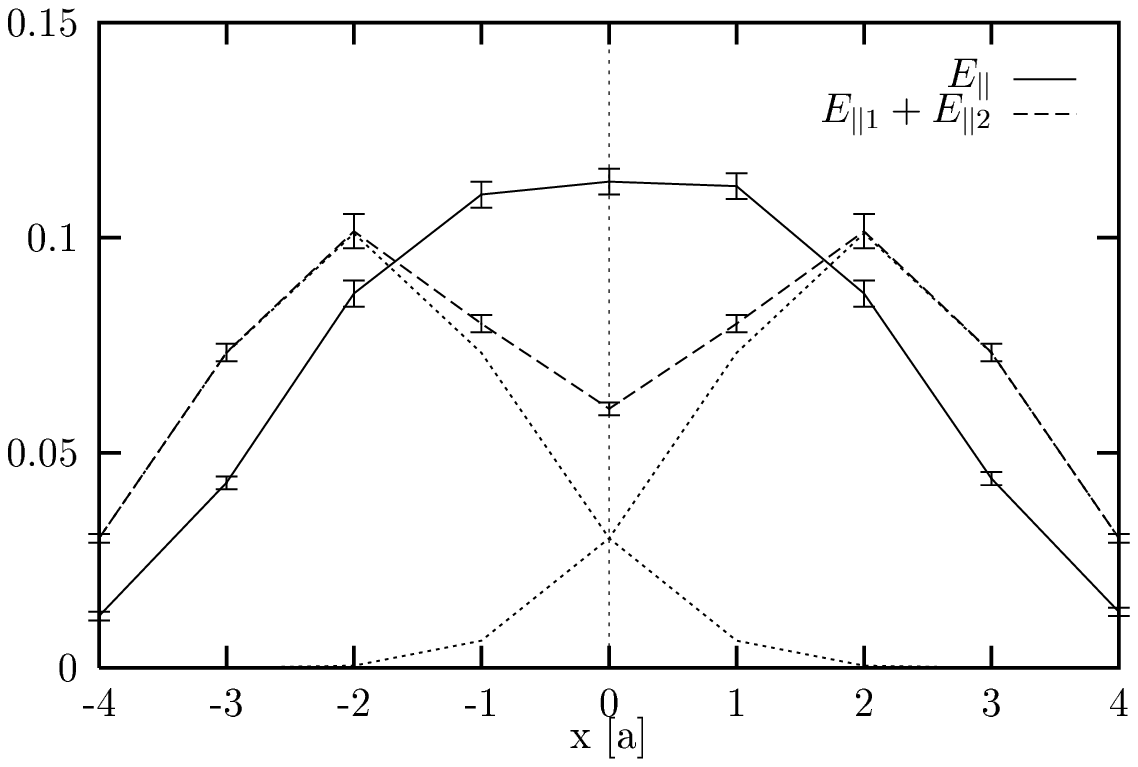}
\end{picture}
\end{center}
\caption[figresone]{
  Results from dual simulations of flux tubes in $U(1)$.
  (a) Ratio of the string tension of a doubly-charged toroidal flux
  tube over that of a singly-charged flux tube with respect to the
  coupling $\beta$. (b) Longitudinal electric field profile of two
  interacting flux tubes in the symmetry plane ($E_{\|}$, solid line)
  for $\beta=0.96$.  The equal charges are separated by $4a$,
  where $a$ is the lattice spacing.  Dotted lines are field profiles
  $E_{\| 1}$ and $E_{\| 2}$ for single flux tubes at $x=-2a$ and $x=+2a$.
  The noninteracting superposition $E_{\| 1}+E_{\| 2}$ is shown as a
  dashed line.
  }
\label{fig:dualsim}
\end{figure}

First, it was shown that the string tension scales proportionally
to the charge rather than the square of the charge in the confinement
phase.  This is shown in Fig.~\ref{fig:dualsim}(a) in which the string
tension of a closed double flux tube is compared to that of a single
closed flux tube as a function of the coupling $\beta$.  These results
were obtained on a $8^3 \times 16$ lattice.  In the Coulomb phase
(large $\beta$ values), the ratio of the string tension of the double
flux tube to that of the single flux tube is 4 in agreement with the
expected result for the energy of a homogeneous field of double strength;
below the transition in the confinement phase, the ratio is 2, in
agreement with the expectations for a dual superconductor.  Flux tube
attraction in the vicinity of the phase transition was confirmed by
an examination of the longitudinal electric field profile at $\beta=0.96$
as shown in Fig.~\ref{fig:dualsim}(b).  Two parallel flux tubes of length
$22a$, where $a$ is the lattice spacing, were brought next to each
other, a transverse distance $4a$ apart, and the resulting longitudinal
electric field in the symmetry plane was measured.  
The profile differed dramatically from
the noninteracting superposition of two single flux tubes; the formation
of one flux tube was observed, in analogy to a type-I superconductor.
These authors also compared their simulation results to the predictions
from a dual London equation.  Agreement was found for small charge distances,
but not for distances greater than ten lattice spacings (they probed
lengths up to $20a$).  They interpreted this as further evidence that the
confining $U(1)$ behaves as an effective type-I superconductor.

\section{Static-quark potentials}

Three groups reported on calculations of static quark-antiquark
potentials; four-quark energies in $SU(2)$ lattice gauge theory
were also presented.

The static-quark potential and its gluon excitations are very
useful probes of confinement.  It is generally believed that at large
quark-antiquark separation $r$, the linearly-growing ground-state
energy of the glue is the manifestation of the confining
flux whose fluctuations can be described in terms of an effective
string theory.  The lowest-lying excitations are then
the Goldstone modes associated with spontaneously-broken transverse
translational symmetry.  Expectations are less clear for small $r$.
The gluon excitations of the static-quark potential are also useful
for studying hybrid heavy-quark mesons using a Born-Oppenheimer expansion.

The first comprehensive determination of the low-lying spectrum of
gluonic excitations in the presence of a static quark-antiquark pair
was presented at this conference\cite{juge}.  The glue energies for
$r$ from 0.1 to 4 fm were extracted from Monte Carlo estimates of
generalized Wilson loops in eight simulations using an improved
gauge-field action.  The use of anisotropic lattices in which the temporal
lattice spacing $a_t$ was much smaller than the spatial spacing $a_s$
was crucial for resolving the glue spectrum, particularly for large $r$.
Finite volume errors were shown to be negligible.  Particular attention
was paid to the volumes for very large $r$:  not only were checks
carried out using additional simulations, but also by studying
volume effects for a naive Nambu-Goto string.  Agreement of energies
obtained using different quark-antiquark orientations on the lattice was
used to check the smallness of finite-spacing errors and to help identify
the continuum rotational quantum numbers corresponding to each level (there
are only discrete symmetries on the lattice).  Results for lattice
spacings ranging from 0.12 to 0.29 fm were obtained to facilitate
extrapolation to the continuum limit. The hadronic scale parameter 
$r_0\approx 0.5$ fm was used to set the scale.

\begin{figure}[t]
\begin{center}
\leavevmode
\epsfxsize=2.9in\epsfbox[78 184 532 678]{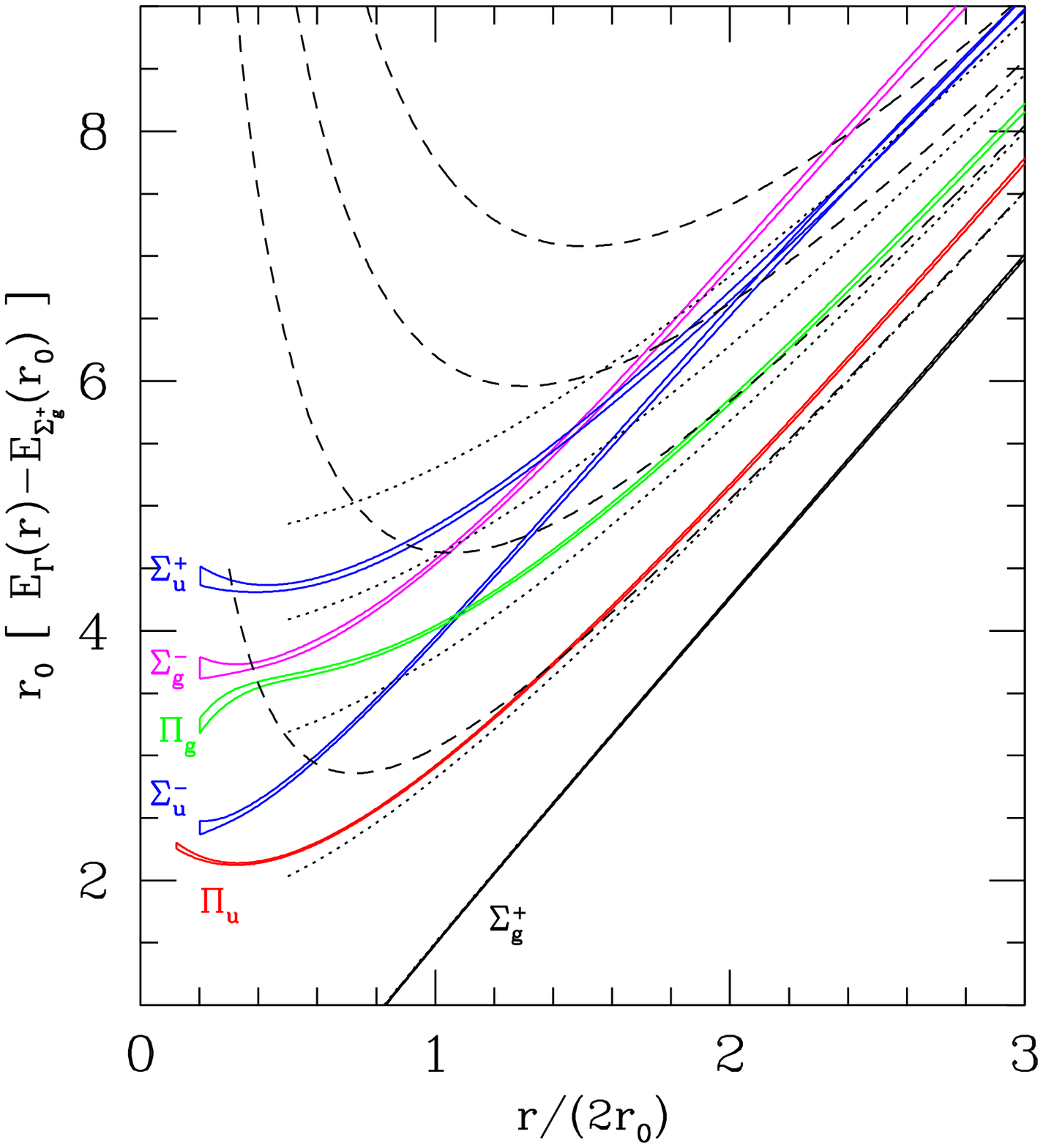}
\epsfxsize=2.9in\epsfbox[78 184 532 678]{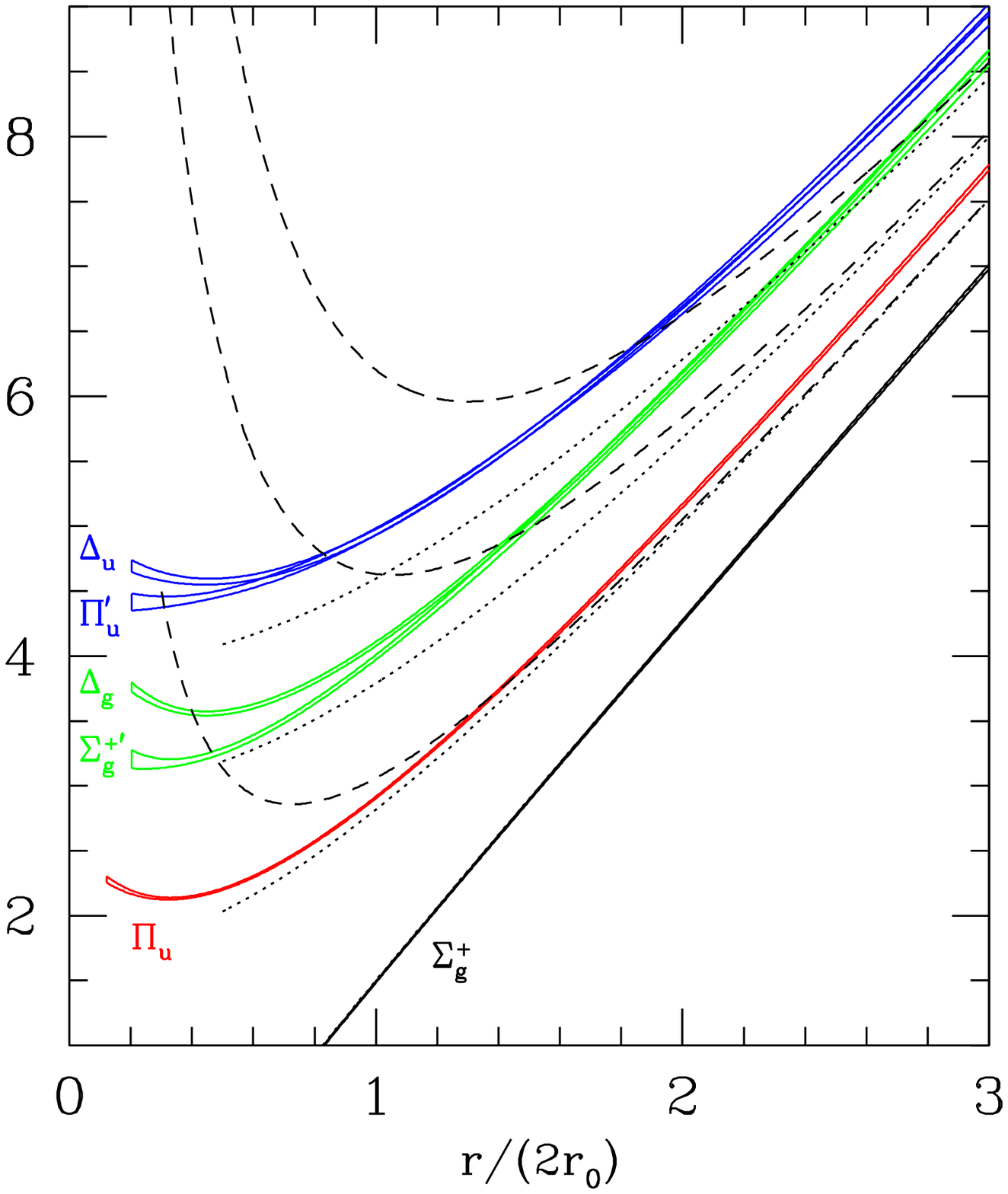}
\end{center}
\caption[figresone]{Gluon excitations of the static-quark potential.
  The dashed lines indicate the locations of
  the $m\pi/r$ gaps above the $\Sigma_g^+$ curve for $m=1$, 2, 3, and 4.
  The dotted curves are the naive Nambu-Goto energies in four-dimensions.
  Note that a $\Phi_u$ interpretation for the
  curve labelled $\Pi_u^\prime$ cannot be ruled out.
  }
\label{fig:res}
\end{figure}
The continuum-limit extrapolations are shown in Fig.~\ref{fig:res}.
The energies are labelled by the magnitude $\Lambda$ of the projection
of the total angular momentum of the gluons onto the molecular axis,
and the symmetry $\eta_{CP}$ under spatial inversion with charge conjugation.
The capital Greek letters $\Sigma, \Pi, \Delta, \Phi, \dots$ are used
to indicate states with $\Lambda=0,1,2,3,\dots$, respectively.
States with $\eta_{CP}=1 (-1)$ are denoted by the subscripts $g$ ($u$).
The $\Sigma$ states are additionally labelled by a superscript $+$ $(-)$
depending on whether they are even (odd) under a reflection in a plane
containing the molecular axis.  The ground-state $\Sigma_g^+$ is the
familiar static-quark potential.  The lowest-lying excitation is the $\Pi_u$.
There is definite evidence of a band structure at large $r$: the 
$\Sigma_g^\prime$, $\Pi_g$, and $\Delta_g$ form the first band above the
$\Pi_u$; the $\Sigma_u^+$, $\Sigma_u^-$, $\Pi_u^\prime/\Phi_u$, and $\Delta_u$
form another band; and the $\Sigma_g^-$ is the highest level.
The level orderings and approximate degeneracies of the gluon energies at
large $r$ match, without exception, those expected of the Goldstone
excitations which are a universal feature of any effective
string theory description of the long confining flux.  However, the precise
$m\pi/r$ gap behaviour is not observed.  For separations less than 2 fm, the
gluon energies lie well below the Goldstone energies and the Goldstone
degeneracies are no longer observed.  The two $\Sigma^-$ states are in
violent disagreement with expectations from a fluctuating string. 
These somewhat surprising results cast serious doubts
on the validity of treating glue in terms of a fluctuating string
for quark-antiquark separations less than 2 fm. 

Some authors\cite{michael,allen} claim that the glue can be described
merely as a mathematical string using the Nambu-Goto action (turning a
blind eye to the fact that the Nambu-Goto string cannot be consistently
quantized in four dimensions).  These results clearly contradict
this claim.  For $r$ greater than 1 fm, the $\Pi_u$ and $\Pi_g$
energies disagree the least with the naive Nambu-Goto energies.  In
all other cases, the discrepancies are significant.  The $\Sigma^-_u$
and $\Sigma_g^-$ levels look nothing like the predictions from the naive
Nambu-Goto string.  The departures of these levels from the Nambu-Goto
energies are so severe that explanations in terms of mixings with
other string modes or consequences of small-$r$ symmetry requirements
are difficult to accept.  Rather, such behaviour signals the complete
failure of the Nambu-Goto intrepretation.

Baker\cite{baker} presented a comparison of results on
the central, spin- and momentum-dependent interquark potentials
from a dual superconductor model with those determined from
first principles using lattice simulations.  The model is an
effective theory of long-distance Yang-Mills in which an octet of
dual potentials are coupled minimally to three octets of scalar Higgs
fields carrying colour magnetic charge.  Monopole condensation
takes place and the dual potentials couple to a quark-antiquark
pair via a Dirac string connecting the pair.  Chromoelectric flux
is confined as the quark-antiquark separation increases as a result
of the dual Meissner effect.  Assuming a particular spontaneous symmetry
breaking sequence, the Lagrangian of the model can be replaced by
an Abelian Higgs Lagrangian.  The model has two parameters, a coupling
constant $\alpha_s$ and the vacuum expectation value of the dual
Higgs field $\phi_0$.  The interquark potentials were calculated
from the Wilson loop of the dual theory.  The two parameters of the
model were determined by fitting to the lattice results for the
central potential.  The remaining nine potentials are then uniquely
determined and were compared to the lattice data.  They were shown
to be in fair agreement for $r$ greater than 0.2 fm; the agreement
was aided by large uncertainties in the simulation results.

Klindworth\cite{klindworth} calculated the static quark potential
in transverse light-front QCD.  In this approach, the two light-cone
coordinates $x^\pm=(x^0\pm x^3)/\sqrt{2}$ are continuous and their
associated degrees of freedom are non-compact, while the other two
transverse directions $x_\perp$ are discretized and the gauge fields
associated with these directions are expressed in terms of compact,
non-unitary link variables.  The calculation was carried out in
the large $N_c$ limit, where $N_c$ is the number of colours.  In
so doing, quark-antiquark pair creation could be ignored.  The
static quark potential was then calculated using Lanczos
diagonalization techniques which necessitated a Fock-space truncation.
A linearly confining potential was found, and an excited potential
was also determined.  The calculation served mainly as an initial
test of this new regularization and computational scheme.  

Furui\cite{fourq} presented results for four-quark energies from
quenched $SU(2)$ simulations.  The aim of the analysis was to
obtain a compact expression for the four-quark interactions
for an arbitrary spatial configuration of the quarks.  They
have examined square, rectangular, tilted rectangular, linear,
quadrilateral, non-planar (at least one link on axis), and
tetrahedral (two links off axis) geometries on a $16^3\times 32$
lattice at two different lattice spacings.  Their results suffer
from significant lattice artifacts.

\section{Conclusion}

A plethora of new results from lattice simulations are shedding
light in some of the previously dark corridors of the labyrinth
surrounding confinement.  Computer simulations of gluons and
quarks remain an important tool for navigating this maze.

\end{document}